# PECULIAR METASTABLE STRUCTURAL STATE IN CARBON STEEL

S.A. Murikov[1], A.V. Shmakov[1], V.N. Urtsev[1], I.L. Yakovleva[2], Yu.N. Gornostyrev[2,3,4],
M.I. Katsnelson[4,5], M.L. Krasnov[5]

[1]*Research and Technological Center Ausferr, Magnitogorsk 455000, Russia*
[2]*Institute of Metal Physics, Ural Division of RAS, Ekaterinburg 620041, Russia*
[3]*Institute of Quantum Materials Science, Ekaterinburg 620075, Russia*
[4]*Ural Federal University, Ekaterinburg 620002, Russia*
[5]*Radboud University Nijmegen, Nijmegen 6525AJ, The Netherlands*
[6]*Magnitogorsk Iron and Steel Works, Magnitogorsk 455000, Russia*

The kinetics of phase transformations at cooling of carbon steel in dependence on the temperature of preliminary annealing $T_{an}$ is studied. It is shown that the cooling from $T_{an} > A_3$ (i.e. above the temperature of ferrite start) with the rate 90 – 100 K/s results in structural state which essentially dependent on $T_{an}$; at $750^0C < T_{an} < 830^0C$ the transformation is of perlite type whereas at $T_{an} > 830^0C$ the martensitic structure arises. Our results evidence the formation of a special structural state in a certain range of temperatures near and above the boundary of two pase region which is characterized by a substantially nanoscale heterogeneity in carbon distribution, lattice distortions, and magnetic short-range order.

1. Introduction

By now steel remains the main construction material of our technology due to high availability of its main components (Fe and C) and diversity of properties reached by realization of various structural states. The latter is possible due to a rich phase diagram of iron with several structural transformations at cooling from high temperatures δ (paramagnetic bcc Fe) → γ (paramagnetic fcc Fe) → α (ferromagnetic bcc Fe). The presence of carbon adds carbide phases, cementite $Fe_3C$ being the most important one [1,2].

Development of the phase transformations in steel includes two main types of processes: the crystal lattice reconstruction and redistribution of carbon between the phases. Competition of these processes results in diversity of structural states of steel which can be realized depending on composition and conditions of thermal treatment. As it commonly accepted (see discussion in Refs [3,4,5]), the variation of magnetic order plays a decisive role in regular evolution of transformation mechanism with temperature from martensite (development of lattice instability at rather deep overcooling) to ferrite (nucleation and growth just below the temperature of γ − α equilibrium). Whereas, the bainite and pearlite transformations take place at intermediate temperature range due to an interplay of diffusion and shear processes [5,6] and results in microstructure with alternating α-Fe and cementite phases.

It is commonly accepted, that the realization of a certain microstructure during thermal treatment depends on the cooling regime, chemical composition [1,2] and the prior austenite grain size [7]. At the same time, the achieved structural state is not sensitive to the start temperature of cooling in austenite region (if other conditions remain the same). Such an assumption is based on the idea that the austenite is a homogeneous solid solution of carbon in fcc iron in whole its phase stability region. This concept is supported by the results of X-ray and neutron diffraction [8] as well as the Mössbauer spectroscopy [9,10] which did not reveal deviations from a random distribution of carbon impurity atoms in fcc iron. As a result, the structural state of the overcooled austenite does not depend on the initial annealing conditions.

In this paper, we demonstrate that this concept is not complete. We show that holding of austenite just above the two-phase γ + α region followed by a rapid cooling results in pearlite transformation while annealing of austenite at higher temperature leads to martensite transformation. The results obtained demonstrate the existence of a peculiar state of austenite in the temperature region close to the boundary of two-phase region what provides a new perspective at the phase transformation in steel.

## 2. Experimental methods

The transformation kinetics was studied by using the specially designed research facility that allows us to measure with high accuracy the temperature changes during cooling/heating and the heat of phase transformation. The samples were made of carbon steel with the base composition 0.65C–0.26Si–0.99Mn (wt%); they were rectangular plates with a thickness of 0.6 mm. To achieve an equilibrium state, the sample was heated in an electric furnace followed by holding for 5 – 10 minutes at temperature $T_{an}$ above the eutectoid point in the range $720^0 – 950^0$C. The sample was then quickly removed from the furnace by a special mechanical device while air cooling system was being activated. The rate of the cooling was ranged by change of pressure before greed of nozzles. The temperature of the sample was measured by three pyrometers operating at different temperature ranges; the following results were obtained by combining indications of all of them. Two pyrometers Raytek Marathon FR1A were operating in the interval of the infrared spectrum corresponding 550-1100°C and the third one was high-speed pyrometer OPTRIS CTfast working in middle infrared interval 50-775°C. One infrared pyrometer was used for calibration, that ensured high accuracy of measurements. The microstructure of samples subjected to thermal treatment and cooling were studied by using SEM QUANTA-200. Investigations of fine features of microstructure were carried out using TEM JEM-200CX.

## 3. Experimental results

As a first step, we investigated the kinetics of transformation in dependence on the cooling rate $V_c$ after exposure of samples at fixed temperature $T_{an} = 900^0$C. In this case the obtained temperature change with time follows the standard concepts [1,2]: for the composition under consideration the diffusive pearlite transformation occurs when the cooling is rather slow and shear martensite colonies gradually replace the pearlite microstructure when the cooling rate increases. The characteristic value of the cooling rate which corresponds to the switching of the transformation mechanism was found at about 100 grad/sec.

The temperature variation during the cooling of samples with initial rate $V_c = 100 - 110$ grad/sec [11] starting from different annealing temperatures $T_{an}$ is shown in Fig. 1. One can see that the temperature profile of the cooling is changed essentially when the temperature $T_{an}$ increases from $720^0$ to $950^0$C. A preliminary exposure of sample just above the two-phase region ($750^0$C $< T_{an} < 830^0$C) leads to a peak in the cooling curves at about $600^0$C. This peak disappears when $T_{an} > 830^0$C and an inflection point is observed on the curves near $240^0$C. The latter feature is typical for the martensitic transformation developing very quickly as a result of a lattice shear instability; the microstructure observed in this case is shown in Fig. 2b. In contrast, the cooling after the exposure at lower temperature, $750^0$C $< T_{an} < 830^0$C, results in the formation of a fine pearlite structure (Fig. 2a) which represents regular alternated plates of α-Fe and cementite $Fe_3C$. It should be noted that the pearlite forms very rapidly in this case, in about 5 sec.

Thus, the observations lead to a surprising conclusion that the mechanism of transformation depends (at least for the steel composition under consideration) on the temperature of pre-annealing in austenite region. For the given cooling rate, the pearlite transformation develops after exposure near the boundary of the austenite region and an increase of the annealing temperature switches the transformation mechanism from pearlitic to martensitic. It should be noted that an increase in the cooling rate above 130 grad/sec (keeping annealing temperature) changes the transformation scenario by replacing pearlite transformation to the martensite one.

Since the used cooling rate was the same for smaller and higher annealing temperature $T_{an}$, it must be assumed that the phase/structural state of the austenite just before the cooling is rather different in these two cases. To qualify the features of the structural state of austenite in the temperature region near the two-phase boundary, an experiment that included the variation of the temperature in the cycle fast heating – cooling (Fig.1 b) starting from the same initial state corresponding to the fine perlite (Fig.2 a) was carried out in Ref. [12]. A characteristic feature of heating curve in this case is well-pronounced plateau near temperature $750^0$C which correspond to the boundary of two-phase region (for the given composition). This plateau is a manifestation of the heat absorption (endothermic reaction) due to dissolution of pearlite colonies. The fast cooling after the fast heating to temperatures above $750^0$C results in pearlite transformation (as is evidenced by the kink on the curves 2-4 in Fig. 1b) or martensite transformation (curve 5 in Fig. 1b).

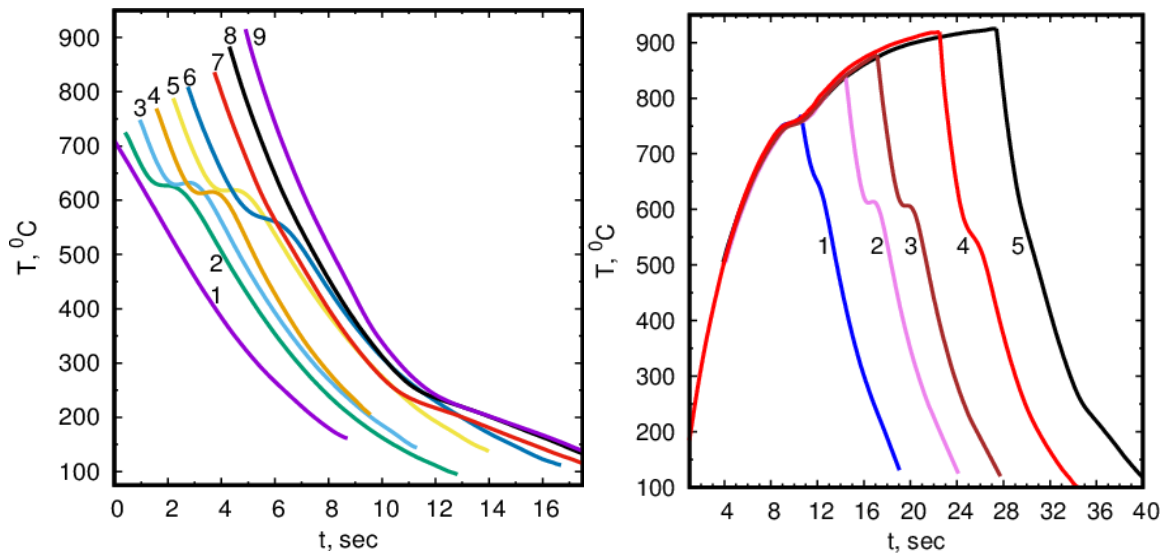

Fig. 1. Temperature of steel samples in dependence on time during fast cooling after annealing above A3 (a) and in cycle fast heating – cooling (b). Curves 1-9 in Fig.1 (a) match annealing temperatures $T_a$ = $720^0$, $750^0$, $770^0$, $790^0$, $810^0$, $830^0$, $850^0$, $900^0$, $950^0$C, respectively. The cooling curves are shifted from each other for convenience of the eye. Note, the temperature start of measurement in Fig.1(a) is slightly lower than corresponding value $T_a$ due to delay caused by movement of the sample from the furnace into the measuring system. Temperatures start of cooling in Fig.1b are equal $765^0$, $840^0$, $875^0$, $920^0$C, $930^0$C (curves 1-5, respectively).

One can assume that the fast heating up to temperature at $750^0$C $< T_{an} < 900^0$C leads to the formation of a microstructure with partially undissolved cementite that initiates perlite transformation during the following fast cooling. However, as one can see from Fig.1a, a preliminary relatively long exposure (5-10 min) in this temperature interval does not change the mechanism of the transformation at the subsequent cooling. These observations clearly point out that a special metastable structural phase/state of austenite that is realized after exposure at temperature region just above two-phase boundary should be different from normal mixture austenite and undissolved pearlite (as assumed in Refs. [12,13]).

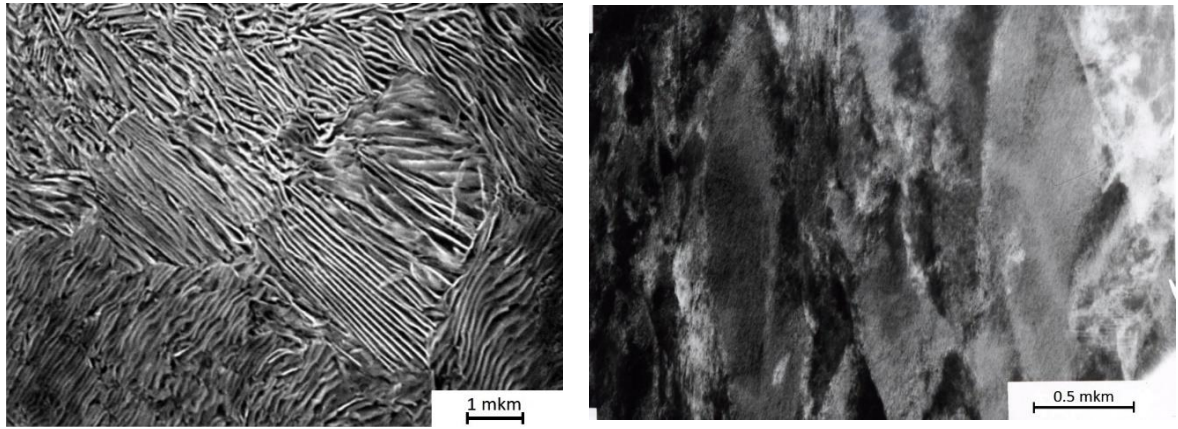

Fig. 2. Microstructure of samples after fast cooling with rate of 110 grad/sec. (a) $T_{an} = 770^0$C, fine plate pearlite; (b) $T_{an} = 850^0$C, martensite.

## 4. Discussion and conclusions

The presented results clearly demonstrate that the scenario of austenite decomposition may be quite complicated and essentially dependent on pre-history of the system in austenite region. This conclusion disagrees with the commonly accepted image of austenite as a homogeneous solid solution of carbon in fcc Fe. Our finding suggests that annealing in a certain range of temperatures near and above the boundary of the austenite region ($750^0$C $< T_{an} < 830^0$C for the composition under consideration) leads to the formation of a special structural state (SSS). The decomposition of this SSS results in the fine pearlite formation (if the cooling rate is not too high) while cooling from a higher temperature (with the same cooling rate) leads to the conventional martensitic transformation. On the other hand, quenching of SSS results in the martensite formation with unusually low tetragonality [13,14]. Based on these observations, the suggestion was made [13,14] that SSS is different from usually assuming homogeneous solid solution of carbon in fcc-Fe and should be considered as a substantially nanoscale heterogeneity in carbon distribution, lattice distortions, and magnetic short-range order (MSRO) [15].

Generally, a short range order corresponding to the low-temperature phase is usually observed in a narrow interval just above the transition temperature. However, the SSS discussed here exists within a rather broad temperature interval. It is known that well-pronounced short-range order or heterogeneity in few or ten nanometers (so called "heterophase fluctuations") are often observed in a certain interval of temperature in magnetic alloys with strong coupling between magnetic and lattice degrees of freedom (or chemical composition), such as Cu – Mn or Fe – Ni invar alloys [16,17]. Usually, the appearance of such internal heterogeneity in varied systems is associated with frustrations which can be of crystallographic origin [18] or result from competition of different interatomic interaction mechanisms [19,20]. The possibility of the formation of some special heterogeneous structural state of Fe-C in austenite region was already briefly discussed in Ref. [13] where it was assigned to a strong coupling between lattice and magnetic degrees of freedom in fcc Fe [21,22].

As follows from the results of *ab initio* calculations the ferromagnetic state of γ-Fe can reduce essentially its energy if its formation is accompanied by tetragonal distortions of crystal lattice [22]; the gain in the magnetic exchange energy is about 0,1 eV (about 1000 K) in this case. It means that, besides the global minimum matching paramagnetic fcc lattice, there is an additional minimum in energy of Fe which corresponds to a metastable magnetically-ordered and tetragonal distorted (fct) state. Wherein, the presence of carbon atoms can increase exchange energy and stabilize fct state [21]. Based on these arguments one should expect the appearance of a heterogeneous SSS of Fe-C with fcc and fct regions alternate each other in some temperature region where MSR is rather strong to provide fct state.

The heterogeneous state discussed here is closely connected to the so called Metastable Intermediate State (MIS) which was recently discussed in detail in Refs. [23,24] based on the results of first principle calculations. According to Ref. [23] MIS can be considered as a tetragonal distorted ferromagnetic austenite with carbon distributed over the interstitial positions. This structure is linked to austenite, ferrite and cementite by natural ways; the use of the concept of MIS allows us to describe correctly austenite decomposition with the pearlite formation [6]. An important condition of the MIS stabilization is the existence of magnetic order (MSRO at least) in the temperature range of interest ($750^0$C - $830^0$C). Based on the results of ab-initio calculations [5,21] we should expect a pronounced increase of the exchange energy in distorted austenite with the moderate carbon concentration and, as follow, stabilization of MIS.

Thus, the concept of a SSS formation in a certain temperature near and above the boundary of the austenite region are supported by current views based on results of ab initio calculations of Fe-Cu system. Further development of such ideas is to be essential for progress in metallurgy technology.

**References**

1. W. C. Leslie and E. Hornbogen, in Physical Metallurgy of Steels, edited by R. W. Cahn and P. Haasen, Physical Metallurgy Vol. 2 (Elsevier, New York, 1996), pp. 1555–1620.
2. R. W. K. Honeycombe and H. K. D. H. Bhadeshia, Steels: Microstructure and Properties, 2nd ed. (Butterworth-Heinemann, Oxford, 1995).
3. L. Kaufman, E. V. Clougherty, R.J. Weiss, Lattice stability of metals-III-iron, Acta Metall., **11** (1963) 323-328
4. H. Hasegawa, D.G. Pettifor, Microscopic Theory of the Temperature-Pressure Phase Diagram of Iron, Phys. Rev. Lett. **50** (1983) 130-133
5. I. K. Razumov, D. V. Boukhvalov, M. V. Petrik, V. N. Urtsev, A. V. Shmakov, M. I. Katsnelson, and Yu. N. Gornostyrev, Role of magnetic degrees of freedom in a scenario of phase transformations in steel, Phys. Rev. B, **90**, (2014) 094101.
6. I. K. Razumov, Yu. N. Gornostyrev, M. I. Katsnelson, Autocatalytic Mechanism of Pearlite Transformation in Steel, Phys. Rev. App., **7**, (2017), 014002
7. M. M. Aranda, B. Kim, R. Rementeria, C. Capdevila, C. García de Andrés, Effect of Prior Austenite Grain Size on Pearlite Transformation in a Hypoeuctectoid Fe-C-Mn Steel, Metallurgical and Materials Transactions A, **45**, (2014) 1778–1786.
8. B. I. Mogutnov, I. A. Tomilin, and L. A. Shvartsman, Thermodynamics of Iron-Carbon Alloys, (Metallurgiya, Moscow, 1972, 328p) [in Russian].
9. W. K. Choo and R. Kaplow, Acta Metall., Mossbauer Measurements on the Aging of Iron-Carbon Martensite, **21**, (1973) 725-732.
10. H. K. D. H. Bhadeshia, Carbon-Carbon Interactions in Iron, J. Mater. Sci. **39**, (2004) 3949-3955.
11. The rate $V_c$ was defined by the slope of the initial part of the cooling curve
12. D. A. Mirzaev, I.L. Yakovleva, N.A. Tereschenko, V.N. Urtsev, V.N. Degtyarev, A.V. Shmakov, Origin of abnormal formation of pearlite in medium-carbon steel under nonequilibrium conditions of heating, Phys. Met. Metallography, **116**, (2016) 572–578
13. V. N. Urtsev, Yu. N. Gornostyrev, M. I. Katsnelson, A. V. Shmakov, A. V. Korolev, V. N. Degtyarev, E. D. Mokshin, and V. I. Voronin, Steel in Translation, **40**, (2010) 671-675.

14. Publication No US-2013-0153090-A1.
15. Early this special structural state of austenite was called as "marinite" [13,14].
16. Y. Tsunoda, N. Orishi, N. Kunitomi, Elastic Moduli of γ-MnCu Alloys, J. Phys. Soc. Japan, **53**, (1984) 359-364.
17. E. F. Wasserman, in Ferromagnetic Materials, edited by K. H. J. Buschow and E. P. Wohlfarth (North-Holland, Amsterdam, 1990), Vol. 5, p. 237.
18. M. Kleman, Curved Crystals, defects and disorder, Adv. Phys., **38**, (1989) 605-667.
19. G. Tarjus, S.A. Kivelson, Z. Nussinov, P. Viot, The frustration based approach of supercooled liquids and the glass transition: a review and critical assessment, J. Phys. : Condens. Matter. **17**, (2005) R1143.
20. I.K. Razumov, Yu. N. Gornostyrev, M.I. Katsnelson, Intrinsic nanoscale inhomogeneity in ordering systems due to elastic-mediated interactions, Europhysics Letters, **80**, (2007) 66001-66005.
21. D. W. Boukhvalov, Yu. N. Gornostyrev, M. I. Katsnelson, and A. I. Lichtenstein, Magnetism and Local Distortions near Carbon Impurity in γ-Iron, Phys. Rev. Lett. **99**, (2007) 247205.
22. S. V. Okatov, A.R. Kuznetsov, Yu. N. Gornostyrev, V.N. Urtsev, M.I. Katsnelson, Effect of magnetic state on the γ-α transition in iron: First-principles calculations of the Bain transformation path, Phys. Rev. B **79**, (2009) 094111.
23. X. Zhang, T. Hickel, J. Rogal, S. Fähler, R. Drautz, and J. Neugebauer, Structural transformations among austenite, ferrite and cementite in Fe-C alloys: A unified theory based on ab initio simulations, Acta Mater. **99**, (2015) 281-289.
24. X. Zhang, T. Hickel, J. Rogal, and J. Neugebauer, Interplay between interstitial displacement and displacive lattice transformations, Phys. Rev. B **94**, (2016) 104109.